\documentclass [12pt,a4paper]{article}
\usepackage{amssymb, theorem}
\newtheorem{thm}{Theorem}[section]
\newtheorem{Thm}{Theorem}[section]

\newtheorem{lemm}{Lemma}[section]
\theorembodyfont{\rm}
\newtheorem{pl}{Example}

\def\prooff{{\it Proof: }}
\def\Cov{\mathrm{Cov}}
\def\Var{\mathrm{Var}}

\def\QCov{\mathrm{Cov}}

\def\qed{\nobreak\hfill $\square$}
\def\<{\langle}
\def\>{\rangle}

\def\fel{{\textstyle{1 \over 2}}}

\def\iA{{\cal A}}
\def\iB{{\cal B}}

\def\bL{{\bf L}}

\def\bM{{\bf M}}
\def\iM{{\cal M}}

\def\im{\mathrm{i}}

\def\bbbn{{\mathbb N}}
\def\bbbr{{\mathbb R}}
\def\bbbc{{\mathbb C}}
\def\Diag{\mbox{Diag}\,}

\def\Tr{\mathrm{Tr}\,}

\def\pont{{\, \cdot \,}}
\def\eps{\varepsilon}

\def\E{{\mathbb E}}
\def\J{{\mathbb J}}

\def\bL{{\mathbb L}}
\def\bR{{ \mathbb R}}

\def\fel{\textstyle{\frac{1}{2}}}
\def\pard{\partial}

\begin{document}
\ \vskip 1cm 
\centerline{\LARGE {\bf Introduction to}}
\bigskip
\centerline{\LARGE {\bf quantum Fisher information}}
\bigskip
\bigskip
\centerline{D\'enes Petz\footnote{E-mail: petz@math.bme.hu.
}}
\centerline{Alfr\'ed R\'enyi Institute of Mathematics,}
\centerline{H-1051 Budapest, Re\'altanoda utca 13-15, Hungary}
\bigskip
\centerline{Catalin Ghinea\footnote{E-mail:  ghinea\_catalin@ceu-budapest.edu}}
\centerline{Department of Mathematics and its Applications}
\centerline{Central European University, 1051 Budapest, N\'ador utca 9, Hungary}

\begin{abstract}
The subject of this paper is a mathematical transition from the Fisher information
of classical statistics to the matrix formalism of quantum theory. If the
monotonicity is the main requirement, then there are several quantum versions
parametrized by a function. In physical applications the minimal is the
most popular. There is a one-to-one correspondence between Fisher informations
(called also monotone metrics) and abstract covariances. The skew information 
and the  $\chi^2$-divergence are treated here as particular cases.
\bigskip

\end{abstract}

\section*{Introduction}

Parameter estimation of probability distributions is one of the most basic
tasks in information theory, and has been generalized to quantum regime
\cite{He, Ho} since the description of quantum measurement is essentially 
probabilistic. First let us have a look at the classical Fisher information. 

Let $(X, \iB, \mu)$ be a probability space. If $\theta=(\theta^1,\dots,
\theta^n)$ is a parameter vector in a neighborhood of $\theta_0 \in \bbbr^n$,
then we should have a smooth family $\mu_\theta$ of probability measures with 
probability density $f_\theta$:
$$
\mu_\theta (H)=\int_H f_\theta(x)\,d\mu(x)\qquad (H \in \iB).
$$
The Fisher information matrix at $\theta_0$ is
\begin{equation}\label{E:1}
J(\mu_\theta ;\theta_0)_{ij}:=\int_X  f_{\theta_0} (x)\frac{\partial}
{\partial \theta^i} \log f_\theta (x)\Big|_{\theta=\theta_0}\,\,
\frac{\partial}{\partial \theta^j} \log f_\theta (x)\Big|_{\theta=\theta_0}
\,d\mu(x)
\end{equation}
\begin{eqnarray*}
&&=\int_X  \frac{1}{f_{\theta_0} (x)}\partial_i f_{\theta_0}(x)
\Big|_{\theta=\theta_0}\,\,
\partial_j f_{\theta_0}(x)\Big|_{\theta=\theta_0}\,d\mu(x)\cr && \cr &&=
-\int_X  f_{\theta_0} (x) \partial_{ij} \log f_\theta (x)
\Big|_{\theta=\theta_0}
\,d\mu(x) \qquad \qquad (1\le i,j \le n).
\end{eqnarray*}
Note that $\log f_\theta (x)$ is usually called {\bf log likelihood} and
its derivative is the {\bf score function}.

The Fisher information matrix is positive semidefinite. For example.
if the parameter $\theta=(\theta^1, \theta^2)$ is two dimensional, then the
Fisher information is a $2 \times 2$ matrix.
From the Schwarz inequality
\begin{eqnarray*}
J(\mu_\theta ; \theta_0)_{12}^2 &\le &\int_X  
\left[\frac{1}{\sqrt{f_{\theta_0} (x)}}
\partial_1 f_{\theta_0}(x)\right]^2\,d\mu(x)
\int_X  \left[\frac{1}{\sqrt{f_{\theta_0} (x)}}
\partial_2 f_{\theta_0}(x)\right]^2\,d\mu(x)\cr && \cr &=&
J(\mu_\theta;\theta_0)_{11} J(\mu_\theta;\theta_0)_{22}
\end{eqnarray*}
Therefore the matrix $J(\mu_\theta;\theta_0)$ is positive semidefinite.

Assume for the sake of simplicity, that $\theta$ is a single parameter.
The random variable $\hat \theta$ is an {\bf unbiased estimator} for the 
parameter $\theta$ if
$$
\E_\theta(\hat \theta):=\int  \hat \theta (x) f_{\theta}(x)\,d\mu(x)=\theta
$$
for all $\theta$. This means that the expectation value of the estimator 
is the parameter. The {\bf Cram\'er-Rao inequality} 
$$
\Var (\hat \theta):=\E_\theta((\hat \theta-\theta)^2)\ge \frac{1}
{J(\mu_\theta;\theta)}
$$
gives a lower bound for the variance of an unbiased estimator. (For more 
parameters we have an inequality between positive matrices.)

In the quantum formalism a probability measure is replaced by a positive matrix
of trace 1. (Its eigenvalues form a probability measure, but to determine
the so-called density matrix a basis of the eigenvectors is also deterministic.)
If a parametrized family of density matrices $D_\theta$ is given, then there
is a possibility for the quantum Fisher information. This quantity is not
unique, the possibilities are determined by linear mappings. The analysis of
the linear mappings is the main issue of the paper. In physics $\theta \in \bbbr$ 
mostly, but if it is an $n$-tuple, then Riemannian geometries appear. A 
coarse-graining gives a monotonicity of the Fisher informations and this is the 
second main subject of the present overview.

Fisher information has a big literature both in the classical and in the quantum case.
The reference of the papers is not at all complete here. The aim is to have an
introduction.
 
\section{A general quantum setting}

The Cram\'er-Rao inequality belongs to the basics of estimation theory
in mathematical statistics. Its quantum analog appeared in the 1970's, 
see the book \cite{He} of Helstrom and the book \cite{Ho} of Holevo. 
Although both the classical Cram\'er-Rao inequality and its quantum 
analog are mathematically as trivial as the Schwarz inequality, the subject 
takes a lot of attention because it is located on the boundary of statistics, 
information and quantum theory. As a starting point we give a very general 
form of the quantum Cram\'er-Rao inequality in the simple setting of finite 
dimensional quantum mechanics. The paper \cite{PD22} is followed here.

For $\theta\in (-\eps, \eps)\subset \bbbr$ a statistical
operator $\rho(\theta)$ is given and the aim is to estimate the value
of the parameter $\theta$ close to $0$. Formally $\rho(\theta)$ is an $n
\times n$ positive semidefinite matrix of trace 1 which describes a
mixed state of a quantum mechanical system and we assume that $\rho(\theta)$
is smooth (in $\theta$). Assume that an estimation
is performed by the measurement of a self-adjoint matrix $A$ playing the
role of an observable. $A$ is called {\bf locally unbiased estimator} if
\begin{equation}\label{E:lue}
\frac{\partial}{\partial \theta}\Tr \rho(\theta) A\Big|_{\theta=0}=1\,.
\end{equation}
This condition holds if $A$ is an {unbiased estimator} for $\theta$,
that is
\begin{equation}
\Tr \rho(\theta) A =\theta \qquad (\theta \in (-\eps,\eps)).
\end{equation}
To require this equality for all values of the parameter is a serious
restriction on the observable $A$ and we prefer to use the weaker
condition (\ref{E:lue}).

Let $[K,L]_{\rho}$ be an inner product (or quadratic cost function) on 
the linear space of self-adjoint matrices. This inner product depends on a density
matrix and its meaning is not described now. When $\rho(\theta)$ is smooth in 
$\theta$, as already was assumed above, then
\begin{equation}\label{E:func}
\frac{\partial}{\partial \theta}\Tr \rho(\theta) B\Big|_{\theta=0}
=[B,L]_{\rho(0)}
\end{equation}
with some $L=L^*$. From (\ref{E:lue}) and (\ref{E:func}), we have
$[A,L]_{\rho(0)}=1$ and the Schwarz inequality yields
\begin{equation}\label{E:CR}
[A,A]_{\rho(0)} \ge \frac{1}{[L,L]_{\rho(0)}}\,.
\end{equation}
This is the celebrated {\bf inequality of Cram\'er-Rao type} for the locally
unbiased estimator. 

The right-hand-side of (\ref{E:CR}) is independent of the estimator
and provides a lower bound for the quadratic cost. The
denominator $[L,L]_{\rho(0)}$ appears to be in the role of Fisher information
here. We call it {\bf quantum Fisher information} with respect to the
cost function $[\pont,\pont]_{\rho(0)}$. This quantity depends on the
tangent of the curve $\rho(\theta)$. If the densities $\rho(\theta)$ and
the estimator $A$ commute, then
$$
L=\rho_0^{-1}\frac{d \rho(\theta)}{d \theta}\quad \mbox{and}\quad
[L,L]_{\rho(0)}=\Tr \rho_0^{-1}\left(\frac{d \rho(\theta)}{d \theta}\right)^2=
\Tr \rho_0\left(\rho_0^{-1}\frac{d \rho(\theta)}{d \theta}\right)^2.
$$
Now we can see some similarity with (\ref{E:1}). 

The quantum Fisher information was defined as $[L,L]_{\rho(0)}$, where
$$
\frac{\partial}{\partial \theta}\rho(\theta)\Big|_{\theta=0}=L.
$$
This $L$ is unique, but the the quantum Fisher information depends on the inner
product $[\pont ,\pont]_{\rho(0)}$. This is not unique, there are several 
possibilities to choose a reasonable inner product $[\pont,\pont]_{\rho(0)}$. Note that
$[A , A]_{\rho(0)}$ should have the interpretation of ``variance'' (if $\Tr
\rho_0 A=0$.)

Another approach is due to Braunstein and Caves \cite{BC} in physics, but
Nagaoka considered a similar approach \cite{Naga}.

\subsection{From classical Fisher information via measurement}

The observable $A$ has a spectral decomposition
$$
A= \sum_{i=1}^k \lambda_i E_i\,.
$$
(Actually the property $E_i^2=E_i$ is not so important, only $E_i \ge 0$ and $\sum_i
E_i=I$. Hence $\{E_i\}$ can be a so-called POVM as well.)
On the set $X=\{1,2,\dots, k\}$ we have probability distributions
$$
\mu_\theta (\{i\})=\Tr \rho(\theta)E_i.
$$
Indeed,
$$
\sum_{i=1}^k \mu_\theta (\{i\})=\Tr \rho(\theta) \sum_{i=1}^k E_i=
\Tr \rho(\theta)=1.
$$
Since
$$
\mu_\theta (\{i\})=\frac{\Tr \rho(\theta)E_i}{\Tr D E_i}\Tr D E_i
$$
we can take
$$
\mu (\{i\})=\Tr D E_i
$$
where $D$ is a statistical operator. Then
\begin{equation} \label{q:3.8}
f_\theta (\{i\})=\frac{\Tr \rho(\theta)E_i}{\Tr D E_i}
\end{equation}
and we have the classical Fisher information defined in ({\ref{E:1}):
$$
\sum_i \frac{\Tr \rho(\theta)E_i}{\Tr D E_i}\left[
\frac{\Tr \rho(\theta)'E_i}{\Tr D E_i}: \frac{\Tr \rho(\theta)E_i}{\Tr D E_i}
\right]^2 \Tr D E_i=
\sum_i \frac{\left[\Tr \rho(\theta)'E_i \right]^2}{\Tr \rho(\theta)E_i}
$$
(This does not depend on $D$.) In the paper \cite{BC} the notation
$$
F(\rho(\theta); \{ E(\xi)\})
=\int \frac{\left[\Tr \rho(\theta)'E(\xi) \right]^2}{\Tr \rho(\theta)E(\xi)}
\, d\xi
$$
is used, this is an integral form, and  for  Braunstein and Caves the quantum 
Fisher information is the supremum of these classical Fisher informations
\cite{BC}.

\begin{Thm}\label{T:1}
Assume that $D$ is a positive definite density matrix, $B=B^*$ and $\Tr B=0$.
If $\rho(\theta)=D+\theta B+ o(\theta^2)$, then the supremum of 
\begin{equation}\label{E:2b}
F(\rho(0); \{E_i\})=\sum_i \frac{\left[\Tr B E_i \right]^2}{\Tr D E_i}
\end{equation}
over the measurements $A= \sum_{i=1}^k \lambda_i E_i$ is
\begin{equation}\label{E:bcf}
\Tr B\J_D^{-1}(B), \qquad \mbox{where} \qquad  \J_D C=(DC+CD)/2.
\end{equation}
\end{Thm}

\prooff
The linear mapping $\J_D$ is invertible, so we can  replace $B$ in (\ref{E:2b}) by 
$\J_D (C)$. We have to show
\begin{eqnarray*}
&&\sum_i \frac{\left[\Tr J_D (C) E_i \right]^2}{\Tr D E_i} \cr &&=
\frac{1}{4}\sum_i \frac{(\Tr CDE_i)^2+ (\Tr DCE_i)^2+2(\Tr CDE_i)(\Tr DCE_i)
}{\Tr D E_i} \le \Tr DC^2.
\end{eqnarray*}
This follows from
\begin{eqnarray*}
\left(\Tr CDE_i\right)^2 &=& \left(\Tr (E_i^{1/2}CD^{1/2})(D^{1/2}E_i^{1/2})\right)^2 
\cr & \le & \Tr E_iCDC \,\, \Tr D^{1/2}E_iD^{1/2} = \Tr E_i CDC \,\, \Tr DE_i. 
\end{eqnarray*}
and
\begin{eqnarray*}
\left(\Tr DCE_i\right)^2 &=& \left(\Tr (E_i^{1/2}D^{1/2})(D^{1/2}CE_i^{1/2})\right)^2 
\cr & \le &\Tr D^{1/2}E_iD^{1/2}\,\, \Tr D^{1/2}CE_iCD^{1/2} = \Tr E_iCDC \,\, \Tr DE_i. 
\end{eqnarray*}
So $F(\rho(0); \{E_i\}) \le \Tr DC^2$ holds for any measurement $\{E_i\}$.

Next we want to analyze the condition for equality. Let $\J_D^{-1}B=C=\sum_k \lambda_k P_k$ 
be the spectral decomposition. In the Scwarz inequalities the condition of equality is
$$
D^{1/2}E_i^{1/2}=c_i D^{1/2}C E_i^{1/2}
$$ 
which is
$$
E_i^{1/2}=c_i C E_i^{1/2}.
$$
So $E_i^{1/2} \le P_{j(i)}$ for a spectral projection $P_{j(i)}$. This implies that all 
projections $P_i$ are the sums of certain $E_i$'s. (The simplest measurement for 
equality corresponds to the observable $C$.) \qed

Note that $\J_D^{-1}$ is in Example \ref{Exe:1}. It is an exercise to show that
for
$$
D=\left[\matrix{r & 0 \cr 0 & 1-r}\right], \qquad
B=\left[\matrix{a & b \cr \overline b & -a}\right]
$$
the optimal observable is
$$
C=\left[\matrix{\displaystyle{\frac{a}{r}} & 2b \cr & \cr 2\overline b & -
\displaystyle{\frac{a}{1-r}}}\right].
$$

The quantum Fisher information (\ref{E:bcf}) is a particular case of the
general approach of the previous session,  $\J_D$ is in Example \ref{Exe:1} below,
this is the minimal quantum Fisher information which is also called {\bf SLD Fisher
information}. The inequality between (\ref{E:2b}) and (\ref{E:bcf}) is a particular 
case of the monotonicity, see \cite{PD2, PD4} and Theorem \ref{T:mon} below.

If $D=\Diag (\lambda_1,\dots, 
\lambda_n)$, then
$$
F_{min}(D; B):=\Tr B\J_D^{-1}(B)=\sum_{ij}\frac{2}{\lambda_i+\lambda_j}|B_{ij}|^2.
$$
In particularly,
$$
F_{min}(D; \im [D,X]) =\sum_{ij}\frac{2(\lambda_i-\lambda_j)^2}{\lambda_i+\lambda_j}
|X_{ij}|^2
$$
and for commuting $D$ and $B$ we have
$$
F_{min}(D; B)=\Tr D^{-1}B^2.
$$

The minimal quantum Fisher information corresponds to the inner product
$$
[A,B]_\rho=\fel \Tr \rho (AB+BA)=\Tr A\J_\rho (B).
$$

Assume now that $\theta=(\theta^1,\theta^2)$. The formula (\ref{q:3.8}) is still true.
If
$$
\partial_i \rho(\theta)=B_i,
$$
then the classical Fisher information matrix $F(\rho(0); \{E_k\})_{ij}$ has the entries
\begin{equation}\label{E:22}
F(\rho(0); \{E_k\})_{ij}=\sum_k \frac{\Tr B_i E_k \Tr B_j E_k}{\Tr \rho(0) E_k}
\end{equation}
and the quantum Fisher information matrix is
\begin{equation}\label{E:22ma}
\left[\matrix{\Tr B_1\J_D^{-1}(B_1) & \Tr B_1\J_D^{-1}(B_2) \cr
\Tr B_2\J_D^{-1}(B_1) & \Tr B_2\J_D^{-1}(B_2)}\right].
\end{equation}
Is there any inequality between the two matrices?

Let $\beta (A)=\sum_k E_k AE_k$. This is a completely positive trace preserving
mapping. In the terminology of Theorem \ref{T:mon2} the matrix (\ref{E:22ma}) is $J_1$ and
$$
J_2=F(\rho(0); \{E_k\}).
$$
The theorem states the inequality $J_2 \le J_1$.

\subsection{The linear mapping $\J_D$}

Let $D \in \bM_n$ be a positive invertible matrix. The linear mapping
$\J_D^f :\bM_n \to \bM_n$ is defined by the formula
$$
\J_D^f=f(\bL_D\bR_D^{-1})\bR_D\,,
$$
where $f: \bbbr^+ \to \bbbr^+$,
$$
\bL_D(X)=D X \qquad\mbox{and}\qquad \bR_D(X)=XD\,.
$$
(The operator $\bL_D\bR_D^{-1}$ appeared in the modular theory of von Neumann
algebras.)

\begin{lemm}
Assume that $f: \bbbr^+ \to \bbbr^+$ is continuous and $D=\Diag(\lambda_1,
\lambda_2,\dots,\lambda_n)$. Then
$$
(\J_D^f B)_{ij}=\lambda_j f\left(\frac{\lambda_i}{\lambda_j}\right)B_{ij}
$$
Moreover, if $f_1 \le f_2$, then $0 \le \J_D^{f_1} \le \J_D^{f_2}$.
\end{lemm}

\prooff
Let $f(x)=x^k$. Then
$$
\J_D^f B= D^k B D^{1-k}
$$
and
$$
(\J_D^f B)_{ij}=\lambda_i^k \lambda_j^{1-k} B_{ij}=\lambda_j f\left(\frac{
\lambda_i}{\lambda_j}\right)B_{ij}.
$$
This is true for polynomials and for any continuous $f$ by approximation. \qed

It follows from the lemma that
\begin{equation}\label{E:212}
\< A, \J_D^f B\>=\< B^*, \J_D^f A^*\>
\end{equation}
if and only if
$$
\lambda_j f\left(\frac{\lambda_i}{\lambda_j}\right)=
\lambda_i f\left(\frac{\lambda_j}{\lambda_i}\right),
$$
which means $xf(x^{-1})=f(x)$. Condition (\ref{E:212}) is equivalent to the
property that $\< X, \J_D^f Y\> \in \bbbr$ when $X$ and $Y$ are self-adjoint.

The functions $f:\bbbr^+ \to \bbbr^+$ 
used here are the {\bf standard operator monotone functions} defined as
\begin{enumerate} 
\item[(i)] if for positive matrices $A \le B$, then $f(A) \le f(B)$,
\item[(ii)] $xf(x^{-1})=f(x)$ and $f(1)=1$.
\end{enumerate}
These functions are between the arithmetic and harmonic means \cite{Ando, pd2}:
$$
\frac{2x}{x+1} \le f(x) \le \frac{1+x}{2}.
$$

Given $f$,
$$
m_f(x,y)=y f\left(\frac{x}{y}\right)
$$
is the corresponding mean and we have
\begin{equation}\label{E:mean}
(\J_D^f B)_{ij}=m_f (\lambda_i, \lambda_j) B_{ij}.
\end{equation}
Hence
$$
\J_D^f B =X \circ B
$$
is a Hadamard product with $X_{ij}=m_f (\lambda_i, \lambda_j)$.
Therefore the linear mapping $\J_D^f$ is positivity preserving if and only 
if the above $X$ is positive. 

The inverse of $\J_D^f$ is the mapping
$$
\frac{1}{f}(\bL_D\bR_D^{-1})\bR_D^{-1} 
$$
which acts as $B \mapsto Y \circ B$ with $Y_{ij}=1/m_f (\lambda_i, \lambda_j)$.
So $(\J_D^f)^{-1}$ is positivity preserving if and only if $Y$ is positive.

A necessary condition for the positivity of $\J_D^f$ is $f(x)\le \sqrt{x}$,
while the necessary condition for the positivity of $(\J_D^f)^{-1}$ is 
$f(x)\ge \sqrt{x}$. So only $f(x)=\sqrt{x}$ is the function which can make 
both mappings positivity preserving. 

\begin{pl}\label{Exe:1}
If $f(x)=(x+1)/2$ (arithmetic mean), then
$$
\J_D B=\frac{1}{2} (DB+BD) \quad\hbox{and}\quad \J_D^{-1} B=\int_0^\infty 
\exp (-tD/2)B \exp (-tD/2)\,dt.
$$
This is from the solution of the equation $DB+BD=2B$. \qed
\end{pl}

\begin{pl}\label{Exe:2}
If $f(x)=2x/(x+1)$ (harmonic mean), then
$$
\J_D B=\int_0^\infty \exp (-tD^{-1}/2)B\exp (-tD^{-1}/2)\,dt 
$$
and
$$
\J_D^{-1} B=\frac{1}{2} (D^{-1}B+BD^{-1}).
$$
This function $f$ is the minimal and it generates the maximal Fisher information
which is also called {\bf right information matrix}. \qed
\end{pl}

\begin{pl}\label{Exe:3}
For the logarithmic mean 
\begin{equation} \label{E:logmean}
f(x)=\frac{x-1}{\log x}
\end{equation}
we have
$$
\J_D(B)=\int_0^1 D^{t}B D^{1-t}\,dt \quad\hbox{and}\quad
\J_D^{-1}(B)=  \int_0^\infty (D+t)^{-1}B(D+t)^{-1}\,dt
$$
This function induces an importan Fisher information. \qed
\end{pl}

\begin{pl}
For the geometric mean $f(x)=\sqrt{x}$ and
$$
\J_D(B)=D^{1/2}B D^{1/2} \quad\hbox{and}\quad
\J_D^{-1}(B)= D^{-1/2}BD^{-1/2}.
$$\qed
\end{pl}

$\J_{D}^f$ is the largest if $f$ is the largest
which is described in Example \ref{Exe:1} and the smallest is in Example \ref{Exe:2}.

\begin{Thm}\label{T:mon}
Let $\beta:\bM_n \to \bM_m$ be a completely positive trace preserving mapping
and $f:\bbbr^+ \to \bbbr^+$ be a matrix monotone function. Then
\begin{equation}\label{E:Fmon}
\beta^* (\J_{\beta(D)}^f)^{-1}\beta  \le (\J_D^f)^{-1}
\end{equation}
and
\begin{equation}\label{E:fimon1}
\beta \J_{D}^f\beta^*  \le \J_{\beta(D)}^f\, .
\end{equation}
\end{Thm}

Actually (\ref{E:Fmon}) and (\ref{E:fimon1}) are equivalent and they are equivalent 
to the matrix monotonicity of $f$ \cite{PD22}. 

In the rest $f$ is always assumed to be a standard matrix monotone function. 
Then $\Tr \J_D B =\Tr DB$.

\begin{pl}
Here we want to study $\J_{D}^f$, when $D$ can have 0 eigenvalues. Formula 
(\ref{E:mean}) makes sense. For example, if $D=\Diag (0, \lambda, \lambda,\mu)$
($\lambda,\mu >0, \lambda \ne \mu$), then
$$
\J_{D}^f B=\left[\matrix{0& m(0,\lambda)B_{12} & m(0, \mu)B_{13}& m(0, \mu)B_{14}
\cr m(0,\lambda)B_{21} & \lambda B_{22} &  m(\lambda, \mu)B_{23}& m(\lambda, \mu)B_{24}
\cr m(0,\mu )B_{31} &  m(\lambda, \mu)B_{32} & \mu B_{34} & \mu B_{34}
\cr m(0,\mu)B_{41} & m(\lambda, \mu)B_{42} &   \mu B_{43} & \mu B_{43} }\right].
$$ 
If $f(0)>0$, then this matrix has only one 0 entry. If $f(0)=0$, then
$$
\J_{D}^f B=\left[\matrix{0& 0 &  0 & 0
\cr 0 & \lambda B_{22} &  m(\lambda, \mu)B_{23}& m(\lambda, \mu)B_{24}
\cr 0 &  m(\lambda, \mu)B_{32} & \mu B_{34} & \mu B_{34}
\cr 0 & m(\lambda, \mu)B_{42} &   \mu B_{43} & \mu B_{43} }\right].
$$ 
and the kernel of $\J_D$ is larger. We have
$$
\< B, \J_{D}^f B\>=\sum_{ij} m_f(\lambda_i,\lambda_j) |B_{ij}|^2
$$
and some terms can be 0 if $D$ is not invertible. 

The inverse of $\J_{D}^f$ exists in the generalized sense
$$
[(\J_{D}^f)^{-1} B]_{ij}=\cases{ \displaystyle{\frac{1}{m_f(\lambda_i,\lambda_j)}B_{ij}}
& if $\quad  m_f(\lambda_i,\lambda_j)\neq 0$,
\cr  \phantom{mm} &\cr 0 & if $\quad  m_f(\lambda_i,\lambda_j) = 0$.}
$$
(This is the Moore-Penrose generalized inverse.)\qed
\end{pl}

It would be interesting to compare the functions which non-zero at 0 with the others.

\section{Fisher information and covariance}

Assume that $f$ is a standard matrix monotone function.
The operators $\J_{D}^f$ are used to define Fisher information and the covariance.
(The latter can be called also quadratic cost.) The operator $\J_{D}^f$ depends
on the function $f$, but $f$ will be not written sometimes. 

Let $A=A^*, B=B^* \in \bM_n$ be observables and $D\in \bM_n$ be a density matrix. 
The covariance of $A$ and $B$  is
\begin{equation}\label{E:Jdef1}
\QCov^f_{D}(A,B):=\< A ,\J_D^f(B)\> -(\Tr DA)(\Tr DB).
\end{equation}
Since
$$
\QCov^f_{D}(A,A) =\< (A-I\Tr DA), \J_D (A-I\Tr DA\>
$$
and $\J_D \ge 0$, we have for the variance $\Var_D^f (A):=\QCov^f_{D}(A,A)  \ge 0$.

The monotonicity (\ref{E:fimon1}) gives
$$
\Var_D^f (\beta^*A) \le \Var_{\beta D}^f(A).
$$
for a completely positive trace preserving mapping $\beta$.

The usual {\bf symmetrized covariance} corresponds to the function $f(t)=(t+1)/2$:
$$
\Cov_D(A,B):=
\frac{1}{2}\Tr (D(A^*B+BA^*))- (\Tr DA^*)(\Tr DB).
$$

Let $A_1,A_2,\dots, A_k$ be self-adjoint matrices and let $D$ be a statistical
operator. The covariance is a $k\times k$ matrix $C(D)$ defined as
\begin{equation}\label{E:CD}
C(D)_{ij}=\QCov^f_{D}(A_i,A_j).
\end{equation}
$C(D)$ is a positive semidefinite matrix and positive definite 
if the observables $A_1,A_2,\dots, A_k$ are linearly independent. It should be 
remarked that this matrix is only a formal analogue of the classical covariance matrix
and it is not related to a single quantum measurement \cite{Luo}.

The variance is defined by $\J_D$ and the Fisher information is formulated by the 
inverse of this mapping:
\begin{equation}\label{E:Jdef2}
\gamma_D(A,B)=\Tr A\J_D^{-1}(B^*).
\end{equation}
Here $A$ and  $B$ are self-adjoint. If $A$ and $B$ are considered as tangent vectors 
at the footpoint $D$, then $\Tr A=\Tr B=0$. In this approach $\gamma_D(A,B)$ is a
an inner product in a Riemannian geometry \cite{AN, H-P}. It seems that this approach
is not popular in quantum theory. It happens also that the condition $\Tr D=1$
is neglected and only $D>0$. Then formula (\ref{E:Jdef2}) can be extended \cite{Wataru}.

If $DA=AD$ for a self-adjoint matrix $A$, then
$$
\gamma_D(A,A)=\Tr D^{-1}A^2
$$
does not depend on the function $f$. (The dependence is characteristic on the orthogonal
complement, this will come later.)

\begin{Thm}\label{T:Fmon}
Assume that $(A,B) \mapsto \gamma_D(A,B)$ is an inner product for $A,B \in \bM_n$, for
positive definite density matrix $D\in \bM_n$ and for every $n$. Suppose the following 
properties:
\begin{enumerate} 
\item[(i)] 
For commuting $D$ and $A=A^*$ we have $\gamma_D(A,A)=\Tr D^{-1}A^2$.
\item[(ii)]
If $\beta:\bM_n \to \bM_m$ is a completely positive trace preserving mapping, then
\begin{equation}\label{E:Fmon2}
\gamma_{\beta(D)}(\beta(A),\beta(A)) \le \gamma_D(A,A).
\end{equation}
\item[(iii)] If $A=A^*$ and $B=B^*$, then $\gamma_D(A,B)$ is a real number.
\item[(iv)] $D\mapsto  \gamma_D(A,B)$ is continuous.
\end{enumerate}
Then
\begin{equation}\label{E:fimon11}
\gamma_D(A,B)=\<A, (\J_{D}^f)^{-1}B\>  
\end{equation}
for a standard matrix monoton function $f$.
\end{Thm}

\begin{pl}
In quantum statistical mechanics, perturbation of a density matrix appears. Suppose
that $D=e^H$ and $A=A^*$ is the perturbation
$$
D_t=\frac{ e^{H+tA}}{\Tr e^{H+tA}} \qquad (t \in \bbbr).
$$
The quantum analog of formula (1) would be
$$
-\Tr D_0 \frac{\partial^2}{\partial t^2} \log D_t\Big|_{t=0}.
$$
A simple computation gives
$$
\int_0^1 \Tr e^{sH}Ae^{(1-s)H}A\,ds- (\Tr DA)^2
$$
This is a kind of variance. \qed
\end{pl}

Let $\iM:=\{D_\theta: \theta \in G\}$ be a smooth $m$-dimensional 
{\bf manifold} of $n \times n$ density matrices. Formally $G \subset 
\bbbr^m$ is an open set including 0. If $\theta \in G$, then $\theta=(\theta_1,
\theta_2, \dots, \theta_m)$. The {\bf Riemannian structure} on $\iM$ is
given by the inner product (\ref{E:Jdef2})
of the tangent vectors $A$ and $B$ at the foot point $D\in \iM$, where
$\J_D: \bM_n \to \bM_n$ is a positive mapping when $\bM_n$ is regarded
as a Hilbert space with the Hilbert-Schmidt inner product. (This means
$\Tr A\J_D(A)^* \ge 0$.)

Assume that a collection $A=(A_1,\dots,A_m)$ of self-adjoint matrices is 
used to estimate the true value of $\theta$. The expectation value of
$A_i$ with respect to the density matrix $D$ is $\Tr DA_i$. $A$ is an 
{\bf unbiased estimator} if
\begin{equation}\label{E:unbi}
\Tr D_\theta A_i=\theta_i \qquad (1 \le i \le n).
\end{equation}
(In many cases unbiased estimator $A=(A_1,\dots,A_m)$ does not exist,
therefore a weaker condition is more useful.)

The {\bf Fisher information} matrix of the estimator $A$ is a positive 
definite matrix 
$$
J(D)_{ij}= \Tr L_i \J_D(L_j), \quad\mbox{where}\quad
L_i = \J_D^{-1}(\partial_i D_\theta).
$$
Both $C(D)$ and $J(D)$ depend on the actual state $D$. 

The next theorem is the the {\bf Cram\'er-Rao inequality} for matrices.
The point is that the right-hand-side does not depend on the estimators.

\begin{Thm}\label{cr-rao} 
Let $A=(A_1,\dots,A_m)$ be an unbiased estimator of $\theta$. Then for the 
above defined matrices the inequality
$$
C(D_\theta) \geq J(D_\theta)^{-1}
$$
holds. 
\end{Thm}

\prooff
In the proof the block-matrix method is used and we restrict 
ourselves for $m=2$ for the sake of simplicity and assume that $\theta=0$. 
Instead of $D_0$ we write $D$.

The matrices $A_1,A_2, L_1, L_2$ are considered as vectors and from the
inner product $\<A,B\>=\Tr A\J_D(B)^*$ we have the positive matrix
$$
X:=\left[\matrix{
\Tr A_1\J_{D }(A_1)&\Tr A_1\J_{D }(A_2)&
\Tr A_1\J_{D }(L_1)&\Tr A_1\J_{D }(L_2)\cr
\Tr A_2\J_{D }(A_1)&\Tr A_2\J_{D }(A_2)&
\Tr A_2\J_{D }(L_1)&\Tr A_2\J_{D }(L_2)\cr
\Tr L_1\J_{D }(A_1)&\Tr L_1\J_{D }(A_2)&
\Tr L_1\J_{D }(L_1)&\Tr L_1\J_{D }(L_2)\cr
\Tr L_2\J_{D }(A_1)&\Tr L_2\J_{D }(A_2)&
\Tr L_2\J_{D }(L_1)&\Tr L_2\J_{D }(L_2)}
\right].
$$  
From the condition (\ref{E:unbi}), we have
$$
\Tr A_i\J_{D }(L_i)=\frac{\partial}{\partial \theta_i}\Tr D_\theta A_i=1 
$$
for $i=1,2$ and
$$
\Tr A_i\J_{D }(L_j)=\frac{\partial}{\partial \theta_j}\Tr D_\theta A_i=0 
$$
if $i \ne j$. Hence the matrix $X$ has the form
\begin{equation}\label{E:kettes}
\left[\matrix{C(D) & I_2 \cr I_2 & J(D)}\right],
\end{equation}
where
$$
C(D)=\left[\matrix{
\Tr A_1\J_{D }(A_1)&\Tr A_1\J_{D }(A_2)\cr
\Tr A_2\J_{D }(A_1)&\Tr A_2\J_{D }(A_2) }\right]
$$
and
$$
J(D)=\left[\matrix{
\Tr L_1\J_{D }(L_1)&\Tr L_1\J_{D }(L_2)\cr
\Tr L_2\J_{D }(L_1)&\Tr L_2\J_{D }(L_2)}\right].
$$
The positivity of (\ref{E:kettes}) implies the statement of the theorem. \qed

We have have the orthogonal decomposition
\begin{equation} \label{E:mero}
\{B=B^* : [D,B]=0\}\oplus \{\im[D,A]:A=A^*\}
\end{equation}
of the self-adjoint matrices and we denote the two subspaces by $\iM_D$ and $\iM_D^c$, 
respectively.

\begin{pl}
The Fisher information and the covariance are easily handled if $D$ is diagonal, 
$D=\Diag (\lambda_1, \dots, \lambda_n)$ or formulated by the matrix units $E(ij)$
$$
D=\sum_i \lambda_i E(ii).
$$
The general formulas in case of diagonal $D$ are
$$
\gamma_D(A,A)=\sum_{ij}\frac{1}{\lambda_jf(\lambda_i/\lambda_j)}|A_{ij}|^2,
\quad
\QCov_{D}(A,A)=\sum_{ij}\lambda_jf(\lambda_i/\lambda_j)|A_{ij}|^2.
$$
Moreover,
\begin{equation}
\gamma_D^f(\im [D,X], \im [D,X] )=\sum_{ij} 
\frac{(\lambda_i -\lambda_j)^2}{\lambda_j f(\lambda_i/\lambda_j)}|X_{ij}|^2.
\end{equation}
Hence for diagonal $D$ all Fisher informations have simple explicit formula.
 
The description of the commutators is more convenient if the eigenvalues are 
different. Let
$$
S_1(ij):=E(ij)+E(ji), \qquad S_2(ij):=-\im E(ij)+\im E(ji)
$$
for $i<j$. (They are the generalization of the Pauli matrices $\sigma_1$ and $\sigma_2$.)
We have
$$
\im [D ,S_1(ij)]= (\lambda_i-\lambda_j)S_2(ij), \qquad
\im [D , S_2(ij)]= (\lambda_j-\lambda_i)S_1(ij).
$$

In Example \ref{Exe:1} we have $f(x)=(1+x)/2$. This gives the minimal Fisher information
described in Theorem \ref{T:1}:
$$
\gamma_D(A,B)=\int_0^\infty \Tr A \exp (-tD/2)B \exp (-tD/2)\,dt.
$$
The corresponding covariance is the symmetrized $\Cov_D(A,B)$. This is maximal among the 
variances.

From Example \ref{Exe:2} we have the maximal Fisher information
$$
\gamma_D(A,B)=\frac{1}{2}\Tr D^{-1}(AB+BA)
$$
The corresponding covariance is a bit similar to the minimal Fisher information:
$$
\QCov_{D}(A,B)=\int_0^\infty \Tr A \exp (-tD^{-1}/2)B \exp (-tD^{-1}/2)\,dt
-\Tr DA\,\Tr DB.
$$

Example \ref{Exe:3} leads to the Boguliubov-Kubo-Mori inner product as Fisher
information \cite{PD3, PD4}:
$$
\gamma_D(A,B)=\int_0^\infty \Tr A(D+t)^{-1}B(D+t)^{-1}\,dt
$$
It is also called BKM Fisher information, the characterization is in the paper \cite{G-S}
and it is also proven that this gives a large deviation bound of consistent superefficient 
estimators \cite{Hayashi0}. \qed
\end{pl}

Let $\iM:=\{\rho(\theta): \theta \in G\}$ be a smooth $k$-dimensional manifold of 
invertible density matrices. The {\it quantum score operators} (or logarithmic 
derivatives) are defined as
\begin{equation}
L_i^f(\theta):= (\J^f_{\rho(\theta)})^{-1}\big(\pard_{\theta_i} \rho(\theta)\big) 
\qquad (1 \le i \le m)
\end{equation}
and
\begin{equation}
J(\theta)_{ij}:=\Tr L_i^f(\theta)\J^f_{\rho(\theta)}\big(L_j(\theta)\big)=
\Tr (\J^f_{\rho(\theta)})^{-1}\big(\pard_{\theta_i} \rho(\theta)\big) (\pard_{\theta_j} 
\rho(\theta))  \qquad (1 \le i,j \le k)
\end{equation}
is the {\bf quantum Fisher information matrix} (depending on the function $f$).
The function $f(x)=(x+1)/2$ yields the symmetric logarithmic derivative (SLD)
Fisher information.

\begin{Thm}\label{T:mon2}
Let $\beta: \bM_n \to \bM_m$ be a completely positive trace preserving mapping
and let $\iM:=\{\rho(\theta)\in \bM_n: \theta \in G\}$ be a smooth $k$-dimensional 
manifold of invertible density matrices. For the Fisher information
matrix $J_1(\theta)$ of $\iM$ and for Fisher information
matrix $J_2(\theta)$ of $\beta(\iM):=\{\beta(\rho(\theta)): \theta \in G\}$ 
we have the monotonicity relation
$$
J_2(\theta) \le J_1(\theta).
$$
\end{Thm}

\prooff
We set $B_i(\theta):= \pard_{\theta_i} \rho(\theta)$. Then
$\J_{\beta(\rho(\theta))}^{-1}\beta(B_i(\theta))$ is the score operator of
$\beta(\iM)$ and we have
\begin{eqnarray*}
\sum_{ij}J_2 (\theta)_{ij} a_i\overline{a_j}&=&
\Tr \J_{\beta(\rho(\theta))}^{-1}\beta \Big(\sum_i a_i B_i(\theta)\Big)
\beta\Big(\sum_j \overline{a_j}B_j(\theta)\Big)\\&=&
\left\<\sum_i a_i B_i,  (\beta^* \J_{\beta)(\rho(\theta))}^{-1}\beta
\sum_j {a_j}B_j(\theta)\right\>\\ &\le&
\left\<\sum_i a_i B_i,   \J_{\rho(\theta)}^{-1}
\sum_j {a_j}B_j(\theta)\right\>\\ & = &
\Tr \J_{\rho(\theta))}^{-1} \Big(\sum_i a_i B_i(\theta)\Big)
\Big(\sum_j \overline{a_j}B_j(\theta)\Big)\\
&=&
\sum_{ij}J_1(\theta)_{ij} a_i\overline{a_j},
\end{eqnarray*}
where (\ref{E:Fmon}) was used. \qed

The monotonicity of the Fisher information matrix in some particular
cases appeared already in the literature: \cite{PD1} treated the case of
the Kubo-Mori inner product and \cite{BC} considered the symmetric
logarithmic derivative and measurement in the role of coarse graining.

\begin{pl}
The function
\begin{equation} \label{E:efek}
f_\beta(t)=\beta(1-\beta)\frac{(x-1)^2}{(x^\beta-1)(x^{1-\beta}-1)}
\end{equation}
is operator monotone if $0 < \beta < 2$. Formally $f(1)$ is not defined, but as
a limit it is 1. The property $xf(x^{-1})=f(x)$ also holds. Therefore this
function determines a Fisher information \cite{PDH}.
If $\beta=1/2$, then the variance has a simple formula:
$$
\Var_D A=\frac{1}{2}\Tr D^{1/2}( D^{1/2} A+AD^{1/2})A- (\Tr DA)^2.
$$\qed
\end{pl}

\begin{pl}
The functions $x^{-\alpha}$ and $x^{\alpha -1}$ are matrix monotone decreasing and so is
their sum. Therefore 
$$
f_\alpha(x) = \frac{2}{x^{-\alpha} + x^{\alpha -1}}
$$
is a standard operator monotone function.
$$
\gamma_\sigma ^{f}(\rho,\rho)= 1+ \Tr \left(\rho-\sigma)\sigma^{-\alpha} 
(\rho-\sigma)\sigma^{\alpha-1} \right)
$$
may remind us to the abstract Fisher information, however now $\rho$ and $\sigma$ are
positive definite density matrices. In the paper \cite{TRus}
$$
\chi^2_\alpha(\rho,\sigma) = \Tr \left(\rho-\sigma)\sigma^{-\alpha} 
(\rho-\sigma)\sigma^{\alpha-1} \right)
$$
is called {\bf quantum $\chi^2$-divergence}. (If $\rho$ and $\sigma$ commute, then the formula 
is independent of $\alpha$. ) Up to the constant 1, this is an interesting and important 
particular case of the monotone metric.  The general theory (\ref{E:Fmon2}) implies
the monotonicity of the  $\chi^2$-divergence. \qed
\end{pl}


\section{Extended monotone metrics}

As an extension of the papers \cite{Cam, PD2} Kuamagai made the following 
generalization \cite{Wataru}. Now $H_n^+$ denotes the strictly positive matrices in 
$\bM_n$. Formally $K_\rho (A,B) \in \bbbc$ is defined for all $\rho\in
H_n^+$, $A,B \in \bM_n$  and $n \in \bbbn$ and it is assumed that
\begin{enumerate} 
\item[(i)] $(A, B) \mapsto K_\rho (A,B)$ is an inner product on $\bM_n$
for every $\rho \in H_n^+$ and $n \in \bbbn$.
\item[(ii)] $\rho \mapsto K_\rho (A,B)$ is continuous.
\item[(iii)] For  a trace-preserving completely positive mapping $\beta$
$$
K_{\beta(\rho)} (\beta(A),\beta(A)) \le  K_\rho (A,A)
$$ 
holds.
\end{enumerate}
In the paper \cite{Wataru} such $K_\rho (A,B)$ is called {\bf extended monotone metric}
and the description is 
$$
K_\rho (A,B)=b(\Tr \rho)\Tr A^* \Tr B+ c \<A, (\J_\rho^f)^{-1} (B)\>,
$$
where $f:\bbbr^+ \to \bbbr^+$ is matrix monotone, $f(1)=1$, $b:\bbbr^+ \to \bbbr^+$
and $c>0$. Note that  
$$
(A, B) \mapsto b(\Tr \rho)\Tr A^* \Tr B\quad \mbox{and} \quad (A, B) \mapsto
c \<A, (\J_\rho^f)^{-1} B\>
$$ 
satisfy conditions (ii) and (iii) with constant $c>0$. The essential point is to check
$$
b(\Tr \rho)\Tr A^*\, \Tr A+ c \<A, (\J_\rho^f)^{-1} A\> \ge 0.
$$
In the case of $1 \times 1$ matrices this is
$$
b(x)|z|^2+ \frac{c}{x}|z|^2 \ge 0
$$
which gives the condition $x b(x)+c >0$. If this is true, then
\begin{eqnarray*}
&&\left( \sum_i \lambda_i \right) b\left( \sum_i \lambda_i \right)
|\sum_i A_{ii}|^2 + c \left( \sum_i \lambda_i \right)
\sum_{ij}\frac{1}{m_f(\lambda_i,\lambda_j)}|A_{ij}|^2 
\cr && \qquad \qquad \ge
-c\Big|\sum_i A_{ii}\Big|^2+c \left( \sum_i \lambda_i \right)
\sum_{ij}\frac{1}{m_f(\lambda_i,\lambda_j)}|A_{ij}|^2 \cr && \qquad \qquad\ge
-c\Big|\sum_i A_{ii}\Big|^2+c \left( \sum_i \lambda_i \right)
\sum_{i}\frac{1}{\lambda_i}|A_{ii}|^2 .
\end{eqnarray*}
The positivity is the inequality
$$
\left( \sum_i \lambda_i \right)\sum_{i}\frac{1}{\lambda_i}|A_{ii}|^2 \ge
\Big|\sum_i A_{ii}\Big|^2
$$
which is a consequence of the Schwarz inequality.

\section{Skew information}

The Wigner-Yanase-Dyson skew information is the quantity
$$
I_p(D,A):= -\frac{1}{2} \Tr [D^p, A][D^{1-p},A] \qquad (0 < p <1).
$$
Actually, the case $p=1/2$ is due to Wigner and Yanase \cite{WYD} 
and the extension was proposed by Dyson. The convexity of $I_p(D,A)$ in 
$A$ is a famous result of Lieb \cite{Lieb}

It was observed in \cite{PDH} that the Wigner-Yanase-Dyson skew information
is connected to the Fisher information which corresponds to the function
(\ref{E:WYD}). For this function we have
\begin{equation} \label{E:WYD}
\gamma_D(\im [D,A],\im [D,A])=\frac{1}{2 \beta(1-\beta)}\Tr\big([\rho^\beta,A]
[\rho^{1-\beta}, A]\big).
\end{equation}
Apart from a constant factor this expression is the skew information
proposed by Wigner and Yanase \cite{WYD}. In the limiting cases $p \to 0$ or $1$ 
we have the function (\ref{E:logmean}) corresponding to the Kubo-Mori-Boguliubov
case.

Let $f$ be a standard function and $A=A^*\in \bM_n$. The quantity
$$
I_D^f(A):= \frac{f(0)}{2}\gamma_D^f(\im [D,A], \im [D,A] )
$$
was called {\bf skew information} in \cite{H} in this general setting. So
the skew information is nothing else but the Fisher information restricted
to $\iM_D^c$, but it is parametrized by the commutator. Skew information appeared 
twenty years before the concept of quantum Fisher information. Skew information 
appears in a rather big literature, for example, connection with uncertainty relations 
\cite{andai, GI, GII, GII:2007a, Kosaki, LZ1, LZ2}.

If $D=\Diag (\lambda_1, \dots, \lambda_n)$ is diagonal, then
$$
\gamma_D^f(\im [D,A], \im [D,A] )=\sum_{ij} 
\frac{(\lambda_i -\lambda_j)^2}{\lambda_j f(\lambda_i/\lambda_j)}|A_{ij}|^2.
$$
This implies that the identity
\begin{equation}\label{E:tilde2}
I_D^f(A)=
\Cov_D(A,A)-\QCov^{\tilde f}_D (A, A )
\end{equation}
holds if $\Tr DA=0$ and
\begin{equation}\label{E:tilde1}
\tilde{f}(x):=\frac{1}{2}\left((x+1)-(x-1)^2 \frac{f(0)}{f(x)}\right).
\end{equation}
It was proved in \cite{3} that for a standard function  $f:\bbbr^+ \to \bbbr$,
$\tilde{f}$ is standard as well. Another proof is in \cite{128skew} which contains 
the following theorem.

\begin{thm}
Assume that $X=X^* \in \iM$ and $\Tr D X=0$. If $f$ is a standard function
such that $f(0)\ne 0$, then
$$
\frac{\pard^2}{\pard t \pard s}  S_F (D+t\im [D,X],D+s\im [D,X])\Big|_{t=s=0}
=f(0)\gamma_D^f(\im [D,X], \im [D,X] )
$$
for the standard function $F=\tilde f$.
\end{thm}

All skew informations are obtained from an $f$-divergence (or quasi-entropy) by
differentiation.

\begin{pl}
The function 
\begin{equation} \label{E:koz}
f(x)=\left(\frac{1+\sqrt{x}}{2}\right)^2
\end{equation}
gives the Wigner-Yanase skew information
$$
I^{WY}(D,A)= I_{1/2}(D,A)= -\frac{1}{2} \Tr [D^{1/2}, A]^2.
$$ 

The skew information coming from the minimal Fisher information and it is 
often denoted as $I^{SLD}(D,A)$. The simple mean inequalities
$$
\left(\frac{1+\sqrt{x}}{2}\right)^2 \le \frac{1+x}{2} 
\le 2 \left(\frac{1+\sqrt{x}}{2}\right)^2
$$
imply
$$
I^{WY}(D,A) \le I^{SLD}(D,A) \le 2 I^{WY}(D,A).
$$
\qed
\end{pl}

\small

\subsection*{Acknowledgement}
This work is supported by the Hungarian Research Grant OTKA 68258.

\end{document}